 \documentclass[final,5p,times,twocolumn]{elsarticle}

 \usepackage{graphics}
 \usepackage{graphicx}
\usepackage{amssymb}
\usepackage{amsmath}

\journal{Journal of Molecular Spectroscopy}

\begin{document}

\begin{frontmatter}

%% Title, authors and addresses

%% use the tnoteref command within \title for footnotes;
%% use the tnotetext command for the associated footnote;
%% use the fnref command within \author or \address for footnotes;
%% use the fntext command for the associated footnote;
%% use the corref command within \author for corresponding author footnotes;
%% use the cortext command for the associated footnote;
%% use the ead command for the email address,
%% and the form \ead[url] for the home page:
%%
%% \title{Title\tnoteref{label1}}
%% \tnotetext[label1]{}
%% \author{Name\corref{cor1}\fnref{label2}}
%% \ead{email address}
%% \ead[url]{home page}
%% \fntext[label2]{}
%% \cortext[cor1]{}
%% \address{Address\fnref{label3}}
%% \fntext[label3]{}

\title{Terahertz spectroscopy of N$^{18}$O and isotopic invariant fit of several 
       nitric oxide isotopologs}

%% use optional labels to link authors explicitly to addresses:
%% \author[label1,label2]{<author name>}
%% \address[label1]{<address>}
%% \address[label2]{<address>}

\author[Koeln]{Holger S.P.~M\"uller\corref{cor1}}
\ead{hspm@ph1.uni-koeln.de}
\cortext[cor1]{Corresponding author.}
\author[Toyama]{Kaori Kobayashi\corref{cor2}}
\ead{kaori@sci.u-toyama.ac.jp}
\cortext[cor2]{Corresponding author.}
\author[Toyama]{Kazumasa Takahashi}
\author[Toyama]{Kazuko Tomaru}
\author[Toyama]{Fusakazu Matsushima}

\address[Koeln]{I.~Physikalisches Institut, Universit{\"a}t zu K{\"o}ln, 
                Z{\"u}lpicher Str. 77, 50937 K{\"o}ln, Germany}
\address[Toyama]{Department of Physics, Faculty of Science, University of Toyama, 
                 3190~Gofuku, Toyama 930-8555, Japan}

%%%%%%%%%%%%%%%%%%%%%%%%%%%%%%%%%%%%%%%%%%%%%%%%%%%%%%%%%%%%%%%%%%%%%%%%%%%%%%%%%%%%%
%%%%%%%%%%%%%%%%%%%%%%%%%%%%%%%%%%%%%%%%%%%%%%%%%%%%%%%%%%%%%%%%%%%%%%%%%%%%%%%%%%%%%
%%%%%%%%%%%%%%%%%%%%%%%%%%%%%%%%%%%%%%%%%%%%%%%%%%%%%%%%%%%%%%%%%%%%%%%%%%%%%%%%%%%%%
\begin{abstract}

A tunable far-infrared laser sideband spectrometer was used to investigate a nitric oxide 
sample enriched in $^{18}$O between 0.99 and 4.75~THz. Regular, electric dipole transitions 
were recorded between 0.99 and 2.52~THz, while magnetic dipole transitions between the 
$^2\Pi_{1/2}$ and $^2\Pi_{3/2}$ spin-ladders were recorded between 3.71 and 4.75~THz. 
These data were combined with lower frequency data of N$^{18}$O (unlabeled atoms refer to 
$^{14}$N and $^{16}$O, respectively), with rotational data of NO, $^{15}$NO, N$^{17}$O, 
and $^{15}$N$^{18}$O, and with heterodyne infrared data of NO to be subjected to one 
isotopic invariant fit. Rotational, fine and hyperfine structure parameters were 
determined along with vibrational, rotational, and Born-Oppenheimer breakdown corrections. 
The resulting spectroscopic parameters permit prediction of rotational spectra suitable 
for the identification of various nitric oxide isotopologs especially in the interstellar 
medium by means of rotational spectroscopy.

\end{abstract}

\begin{keyword}  %%% up to 6 !!
%% keywords here, in the form: keyword \sep keyword

nitric oxide \sep
terahertz spectroscopy \sep 
electric dipole transitions \sep
magnetic dipole transitions \sep
fine structure \sep
hyperfine structure

%% MSC codes here, in the form: \MSC code \sep code
%% or \MSC[2008] code \sep code (2000 is the default)

\end{keyword}

\end{frontmatter}

%%
%% Start line numbering here if you want
%%
% \linenumbers

%%%%%%%%%%%%%%%%%%%%%%%%%%%%%%%%%%%%%%%%%%%%%%%%%%%%%%%%%%%%%%%%%%%%%%%%%%%%%%%%%%%%%
%%%%%  main text  %%%%%%%%%%%%%%%%%%%%%%%%%%%%%%%%%%%%%%%%%%%%%%%%%%%%%%%%%%%%%%%%%%%
%%%%%%%%%%%%%%%%%%%%%%%%%%%%%%%%%%%%%%%%%%%%%%%%%%%%%%%%%%%%%%%%%%%%%%%%%%%%%%%%%%%%%

%%%%%%%%%%%%%%%%%%%%%%%%%%%%%%%%%%%%%%%%%%%%%%%%%%%%%%%%%%%%%%%%%%%%%%%%%%%%%%%%%%%%%
%%%%%  Introduction  %%%%%%%%%%%%%%%%%%%%%%%%%%%%%%%%%%%%%%%%%%%%%%%%%%%%%%%%%%%%%%%%
%%%%%%%%%%%%%%%%%%%%%%%%%%%%%%%%%%%%%%%%%%%%%%%%%%%%%%%%%%%%%%%%%%%%%%%%%%%%%%%%%%%%%

\section{Introduction}
\label{introduction}

Nitric oxide, NO, is the only stable diatomic molecule with an odd number of electrons. 
It is, therefore, of great interest for fundamental sciences and in particular for 
molecular spectroscopy. This is an important reason for a large body of spectroscopic 
investigations into the ground $^2\Pi_{\rm r}$ electronic state of NO. 
Soon after an electron paramagnetic resonance study of NO in 1950 \cite{NO_ESR_1950}, 
the first reports on its rotational spectrum in the ground vibrational state appeared 
\cite{NO_rot_1953,NO_rot_1956,NO_rot_1959}. Further studies were carried out later on 
the main isotopolog 
\cite{NO_rot_1979,NO_rot_1980,NO_N-15-O_NO-18_rot_1991,NO_N-15-O_rot_1999,NO_N-15-O_NO-18_rot_1999}, 
on $^{15}$NO \cite{NO_N-15-O_NO-18_rot_1991,NO_N-15-O_rot_1999,NO_N-15-O_NO-18_rot_1999}, 
on N$^{18}$O \cite{NO_N-15-O_NO-18_rot_1991,NO_N-15-O_NO-18_rot_1999}, and even 
on N$^{17}$O and $^{15}$N$^{18}$O \cite{NO-17_N-15-O-18_rot_1994}; unlabeled atoms refer 
to $^{14}$N and $^{16}$O. The $\Lambda$-doubling transitions in the radio-frequency (RF) 
and microwave (MW) regions were studied extensively for NO and $^{15}$NO in their ground 
vibrational states \cite{NO_N-15-O_RF_1972} with additional data for NO 
\cite{NO_RF_1970,NO_RF_1976,NO_v=0+1_IR-RF-DR_1977,NO_v=0+1_IR-RF-DR_1981}, even in its 
excited vibrational state $\varv = 1$ \cite{NO_v=0+1_IR-RF-DR_1977,NO_v=0+1_IR-RF-DR_1981}. 
The spin-orbit splitting in $\varv = 0$ was determined directly from high-resolution 
observations of the $^2\Pi_{3/2}$~$\leftarrow$~$^2\Pi_{1/2}$ magnetic dipole spectrum 
of NO \cite{NO_N-15-O_rot_1999,NO_magnetic_1992} and of $^{15}$NO \cite{NO_N-15-O_rot_1999} 
near 123~cm$^{-1}$. Numerous infrared (IR) studies have been carried out, mostly on the 
main isotopic species. Among those with experimental transitions frequencies we mention 
in particular heterodyne \cite{NO_heterdyne-IR_1986} and Lamb-dip heterodyne studies 
\cite{NO_heterdyne-IR_1996} of the fundamental vibrational band, which was also 
recorded with Fourier transform spectroscopy (FTS) 
\cite{NO_IR_1-0_1994,NO_IR_1-0-forbidden_1994,NO_IR_1-0_1995}. 
Much higher vibrational  levels were accessed through emission spectroscopy of the 
first \cite{NO_IR-emi_2-0_to_6-4_1979,NO_IR_emi_2-0_to_15-13+_1980} and second 
overtone \cite{NO_IR-emi_10-7_to_22-19_1982}. Isotopic data are also available, 
albeit to a lesser extent \cite{565758_IR_1-0_+_1979,5658_IR_1_2_3-0_1980}.

The spectroscopy of NO is also important for diagnostic purposes. Nitric oxide is a 
minor constituent of Earth's atmosphere with a prominent role in the catalytic 
decomposition of ozone in the stratosphere. However, it is not easily detected in 
the atmosphere employing rotational sepctroscopy because of its small dipole 
moment of 0.15872\,(2)~D \cite{NO_RF_1970}, high-resolution IR spectroscopy of 
its fundamental band is commonly used instead. In fact, we are only aware of one 
report on microwave observations of atmospheric NO \cite{NO_atmo_rot_1992}. 
The NRAO 11~m telescope on Kitt Peak was used to to record the $J = 2.5^f - 1.5^f$ 
transition of NO near 250.8~GHz. Filtering out emission with line widths larger than 
1.5~MHz, they were sensitive only to NO at high altitudes of $> 50$~km. 
Radio astronomy, on the other hand, was used frequently to observe NO in space. 
Nitric oxide was detected in the star-forming region Sagittarius B2(OH) \cite{NO_det_1978} 
and in dark clouds \cite{NO_dark-cloud_1990}. Its abundance in these dense molecular 
clouds is rather high, around that of C$^{18}$O \cite{NO_abundance_1992}, so it was hardly 
surprising that it was also detected in external galaxies such as the star-burst galaxy 
NGC~253 \cite{NO_extragal_2003}. Higher rotationally excited transitions of NO have been 
observed with the high-resolution instrument HIFI on board of the \textit{Herschel} 
space observatory in the frame work of molecular line surveys of the prolific 
star-forming regions Orion~KL \cite{Orion-KL_HIFI_2014} and Sagittarius B2(N) 
\cite{Sgr-B2(N)_HIFI_2014}. The Atacama Large Millimeter Array (ALMA), which is 
currently under construction, will provide not only very high spatial resolution, 
but also very high sensitivity and spectral resolution, which should facilitate 
the detection of minor isotopic species of NO.

We recorded magnetic dipole transitions of N$^{18}$O around 4~THz to determine the 
spin-orbit splitting directly. In addition, we recorded electric dipole transitions 
around 2~THz for better prediction of higher rotational states. The resulting data 
were combined with other N$^{18}$O rotational data to determine its spectroscopic 
parameters. Ultimately, the data were also combined with rotational data of other 
NO isotopologs and with heterodyne IR data of the main species for an isotopic 
invariant fit along with Born-Oppenheimer breakdown (BOB) corrections to derive, 
in turn, predictions of the rotational spectra of NO isotopic species for radio 
astronomy.

%%%%%%%%%%%%%%%%%%%%%%%%%%%%%%%%%%%%%%%%%%%%%%%%%%%%%%%%%%%%%%%%%%%%%%%%%%%%%%%%%%%%%
%%%%%  Experimental details and observed spectrum  %%%%%%%%%%%%%%%%%%%%%%%%%%%%%%%%%%
%%%%%%%%%%%%%%%%%%%%%%%%%%%%%%%%%%%%%%%%%%%%%%%%%%%%%%%%%%%%%%%%%%%%%%%%%%%%%%%%%%%%%

\section{Experimental details and observed spectrum}
\label{exptl_details}

The terahertz spectrometer used at the University of Toyama in the present study 
is a so-called Evenson-type tunable far-infrared spectrometer (TuFIR) based on 
a frequency synthesizing technique developed by Evenson and co-workers \cite{exp_1}. 
Details of the spectrometer can be found elsewhere \cite{exp_2}. The basic principle 
is the stable far-infrared genaration by the difference frequency generated from 
two frequency-stabilized CO$_2$ lasers. The difference frequency is mixed with 
the microwave radiation from a synthesized sweeper on a metal-insulator-metal 
(MIM) diode to achieve tunability. Two side bands (upper and lower) are generated. 
The frequency of the absorption can be determined by the phase of the signal.  
A liquid-helium-cooled Si bolometer is used to detect the terahertz radiation.  
The 1$f$ detection signal from the lock-in amplifier was recorded with a computer. 
A path length modulator was inserted into the terahertz path in order to eliminate 
standing waves.

Two glass cells, 250~cm or 40~cm long, were used for most of the measurements.  
The isotopically enriched N$^{18}$O (Shoko Co. Ltd., 97\,\% $^{18}$O) was 
used without further purification. The sample pressure was maintained at about 
7$-$12~Pa for the pure rotational (electronic dipole) transitions and at $\sim$120~Pa 
for the weak magnetic transitions. All measurements were carried out at room temperature.

%%%%%%%%%%%%%%%%%%%%%%%%%%%%%%%%%%%%%%%%%%%%%%%%%%%%%%%%%%%%%%%%%%%%%%%%%%%%%%%%%%%%%
%%%%%  Figure 3  %%%%%%%%%%%%%%%%%%%%%%%%%%%%%%%%%%%%%%%%%%%%%%%%%%%%%%%%%%%%%%%%%%%%
%%%%%%%%%%%%%%%%%%%%%%%%%%%%%%%%%%%%%%%%%%%%%%%%%%%%%%%%%%%%%%%%%%%%%%%%%%%%%%%%%%%%%

\begin{figure}
  \includegraphics[width=8.8cm]{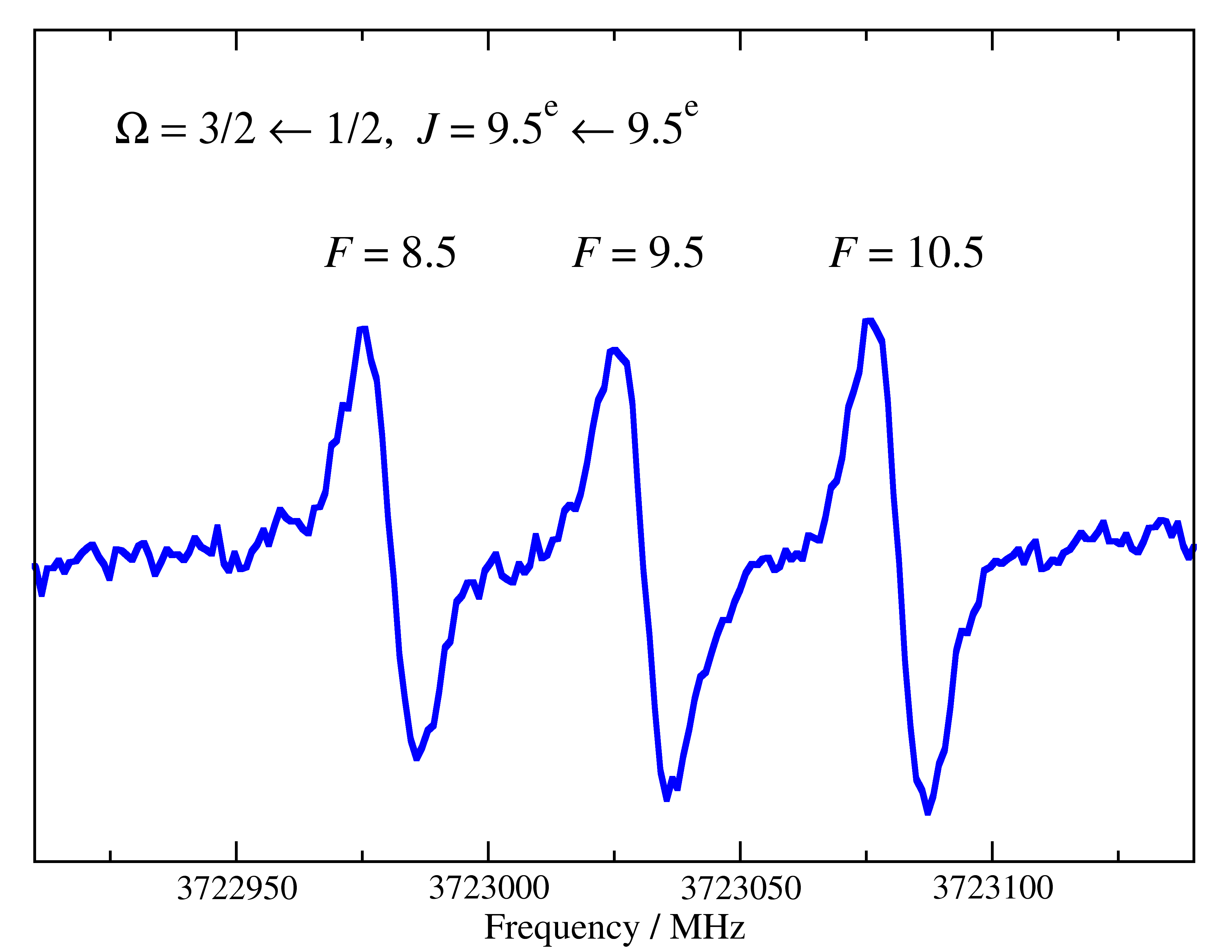}
  \caption{Terahertz spectrum of N$^{18}$O in the region of the 
           $^2\Pi_{3/2}$~$\leftarrow$~$^2\Pi_{1/2}$, $Q(9.5^{\rm e})$ fine structure 
           transition with resolved $^{14}$N hyperfine splitting.}
  \label{FS-spectrum}
\end{figure}

%%%%%%%%%%%%%%%%%%%%%%%%%%%%%%%%%%%%%%%%%%%%%%%%%%%%%%%%%%%%%%%%%%%%%%%%%%%%%%%%%%%%%
%%%%%  Table 1  %%%%%%%%%%%%%%%%%%%%%%%%%%%%%%%%%%%%%%%%%%%%%%%%%%%%%%%%%%%%%%%%%%%%%
%%%%%%%%%%%%%%%%%%%%%%%%%%%%%%%%%%%%%%%%%%%%%%%%%%%%%%%%%%%%%%%%%%%%%%%%%%%%%%%%%%%%%

\begin{table}
  \begin{center}
  \caption{Rotational (electric dipole) transitions of N$^{18}$O, frequency$^a$ (MHz) and 
           residuals O$-$C (MHz) between observed frequency and that calculated from the 
           isotopic invariant fit.}
  \label{rotational_transitions}
\smallskip
{\footnotesize
  \begin{tabular}{ccr@{}lr@{}l}
  \hline
$\Omega$ & $J'^{\rm p'} - J''^{\rm p''}$      & \multicolumn{2}{c}{Frequency} & \multicolumn{2}{c}{O$-$C} \\[1pt]
\hline
$^2\Pi_{1/2}$ & $10.5^{\rm e} -  9.5^{\rm e}$ &  998525&.402 &    0&.037 \\
$^2\Pi_{1/2}$ & $10.5^{\rm f} -  9.5^{\rm f}$ &  998824&.023 &    0&.061 \\
$^2\Pi_{3/2}$ & $10.5^{\rm f} -  9.5^{\rm f}$ & 1024876&.524 & $-$0&.074 \\
$^2\Pi_{1/2}$ & $11.5^{\rm e} - 10.5^{\rm e}$ & 1093603&.395 &    0&.012 \\
$^2\Pi_{1/2}$ & $11.5^{\rm f} - 10.5^{\rm f}$ & 1093894&.752 &    0&.061 \\
$^2\Pi_{3/2}$ & $11.5^{\rm e} - 10.5^{\rm e}$ & 1122177&.958 &    0&.027 \\
$^2\Pi_{3/2}$ & $11.5^{\rm f} - 10.5^{\rm f}$ & 1122224&.596 &    0&.065 \\
$^2\Pi_{1/2}$ & $12.5^{\rm e} - 11.5^{\rm e}$ & 1188668&.423 &    0&.017 \\
$^2\Pi_{1/2}$ & $12.5^{\rm f} - 11.5^{\rm f}$ & 1188952&.059 &    0&.026 \\
$^2\Pi_{3/2}$ & $12.5^{\rm e} - 11.5^{\rm e}$ & 1219450&.321 &    0&.074 \\
$^2\Pi_{3/2}$ & $12.5^{\rm f} - 11.5^{\rm f}$ & 1219504&.465 &    0&.012 \\
$^2\Pi_{1/2}$ & $13.5^{\rm e} - 12.5^{\rm e}$ & 1283718&.135 &    0&.012 \\
$^2\Pi_{1/2}$ & $13.5^{\rm f} - 12.5^{\rm f}$ & 1283993&.739 &    0&.024 \\
$^2\Pi_{3/2}$ & $13.5^{\rm e} - 12.5^{\rm e}$ & 1316649&.397 &    0&.027 \\
$^2\Pi_{3/2}$ & $13.5^{\rm f} - 12.5^{\rm f}$ & 1316711&.601 &    0&.044 \\
$^2\Pi_{1/2}$ & $14.5^{\rm e} - 13.5^{\rm e}$ & 1378750&.049 &    0&.040 \\
$^2\Pi_{1/2}$ & $14.5^{\rm f} - 13.5^{\rm f}$ & 1379017&.279 &    0&.026 \\
$^2\Pi_{3/2}$ & $14.5^{\rm e} - 13.5^{\rm e}$ & 1413770&.721 & $-$0&.029 \\
$^2\Pi_{3/2}$ & $14.5^{\rm f} - 13.5^{\rm f}$ & 1413841&.225 & $-$0&.019 \\
$^2\Pi_{1/2}$ & $15.5^{\rm e} - 14.5^{\rm e}$ & 1473761&.367 &    0&.031 \\
$^2\Pi_{1/2}$ & $15.5^{\rm f} - 14.5^{\rm f}$ & 1474020&.025 &    0&.065 \\
$^2\Pi_{3/2}$ & $15.5^{\rm e} - 14.5^{\rm e}$ & 1510810&.030 & $-$0&.008 \\
$^2\Pi_{3/2}$ & $15.5^{\rm f} - 14.5^{\rm f}$ & 1510889&.084 & $-$0&.039 \\
$^2\Pi_{1/2}$ & $16.5^{\rm e} - 15.5^{\rm e}$ & 1568749&.227 &    0&.051 \\
$^2\Pi_{1/2}$ & $16.5^{\rm f} - 15.5^{\rm f}$ & 1568998&.950 &    0&.001 \\
$^2\Pi_{1/2}$ & $19.5^{\rm e} - 18.5^{\rm e}$ & 1853539&.862 &    0&.038 \\
$^2\Pi_{1/2}$ & $19.5^{\rm f} - 18.5^{\rm f}$ & 1853761&.992 & $-$0&.033 \\
$^2\Pi_{3/2}$ & $19.5^{\rm f} - 18.5^{\rm f}$ & 1898181&.516 &    0&.036 \\
$^2\Pi_{1/2}$ & $26.5^{\rm e} - 25.5^{\rm e}$ & 2516576&.862 & $-$0&.009 \\
$^2\Pi_{1/2}$ & $26.5^{\rm f} - 25.5^{\rm f}$ & 2516732&.584 & $-$0&.046 \\
\hline \hline
\end{tabular}\\
}
\end{center}
$^a$ {\footnotesize
Uncertainty 50~kHz for each rotational line.}
\end{table}

%%%%%%%%%%%%%%%%%%%%%%%%%%%%%%%%%%%%%%%%%%%%%%%%%%%%%%%%%%%%%%%%%%%%%%%%%%%%%%%%%%%%%
%%%%%  Table 2  %%%%%%%%%%%%%%%%%%%%%%%%%%%%%%%%%%%%%%%%%%%%%%%%%%%%%%%%%%%%%%%%%%%%%
%%%%%%%%%%%%%%%%%%%%%%%%%%%%%%%%%%%%%%%%%%%%%%%%%%%%%%%%%%%%%%%%%%%%%%%%%%%%%%%%%%%%%

\begin{table}
  \begin{center}
  \caption{$^2\Pi_{3/2}$~$\leftarrow$~$^2\Pi_{1/2}$ magnetic dipole (fine structure) transitions 
           of N$^{18}$O, frequency$^a$ (MHz) and residuals O$-$C (MHz) between observed frequency 
           and that calculated from the isotopic invariant fit.}
  \label{FS_transitions}
\smallskip
{\footnotesize
  \begin{tabular}{ccr@{}lr@{}l}
  \hline
$J'^{\rm p'} - J''^{\rm p''}$ & $F' - F''$ & \multicolumn{2}{c}{Frequency} & \multicolumn{2}{c}{O$-$C} \\[1pt]
\hline
 $ 9.5^{\rm f} -  9.5^{\rm f}$ & 10.5 $-$ 10.5 & 3719850&.891 & $-$0&.225 \\
 $ 9.5^{\rm f} -  9.5^{\rm f}$ &  9.5 $-$  9.5 & 3719910&.708 & $-$0&.097 \\
 $ 9.5^{\rm f} -  9.5^{\rm f}$ &  8.5 $-$  8.5 & 3719968&.880 & $-$0&.152 \\
 $ 9.5^{\rm e} -  9.5^{\rm e}$ &  8.5 $-$  8.5 & 3722980&.772 &    0&.047 \\
 $ 9.5^{\rm e} -  9.5^{\rm e}$ &  9.5 $-$  9.5 & 3723030&.929 &    0&.052 \\
 $ 9.5^{\rm e} -  9.5^{\rm e}$ & 10.5 $-$ 10.5 & 3723081&.461 & $-$0&.027 \\
 $10.5^{\rm f} - 10.5^{\rm f}$ & 11.5 $-$ 11.5 & 3745903&.024 &    0&.302 \\
 $10.5^{\rm f} - 10.5^{\rm f}$ & 10.5 $-$ 10.5 & 3745963&.357 &    0&.121 \\
 $10.5^{\rm f} - 10.5^{\rm f}$ &  9.5 $-$  9.5 & 3746023&.090 & $-$0&.049 \\
 $10.5^{\rm e} - 10.5^{\rm e}$ &  9.5 $-$  9.5 & 3749294&.640 &    0&.089 \\
 $10.5^{\rm e} - 10.5^{\rm e}$ & 10.5 $-$ 10.5 & 3749343&.231 &    0&.055 \\
 $10.5^{\rm e} - 10.5^{\rm e}$ & 11.5 $-$ 11.5 & 3749391&.245 &    0&.028 \\
 $ 4.5^{\rm e} -  3.5^{\rm f}$ &  5.5 $-$  4.5 & 4053636&.418 & $-$0&.057 \\
 $ 4.5^{\rm e} -  3.5^{\rm f}$ &  4.5 $-$  3.5 & 4053694&.601 &    0&.000 \\
 $ 4.5^{\rm e} -  3.5^{\rm f}$ &  3.5 $-$  2.5 & 4053738&.502 &    0&.054 \\
 $ 4.5^{\rm f} -  3.5^{\rm e}$ &  3.5 $-$  2.5 & 4054966&.173 & $-$0&.226 \\
 $ 4.5^{\rm f} -  3.5^{\rm e}$ &  4.5 $-$  3.5 & 4055021&.910 & $-$0&.696 \\
 $ 4.5^{\rm f} -  3.5^{\rm e}$ &  5.5 $-$  4.5 & 4055093&.835 &    0&.279 \\
 $ 6.5^{\rm e} -  5.5^{\rm f}$ &  7.5 $-$  6.5 & 4274260&.709 &    0&.446 \\
 $ 6.5^{\rm e} -  5.5^{\rm f}$ &  6.5 $-$  5.5 & 4274321&.120 &    0&.416 \\
 $ 6.5^{\rm e} -  5.5^{\rm f}$ &  5.5 $-$  4.5 & 4274371&.189 & $-$0&.056 \\
 $ 6.5^{\rm f} -  5.5^{\rm e}$ &  5.5 $-$  4.5 & 4276288&.069 &    0&.396 \\
 $ 6.5^{\rm f} -  5.5^{\rm e}$ &  6.5 $-$  5.5 & 4276341&.083 & $-$0&.024 \\
 $ 6.5^{\rm f} -  5.5^{\rm e}$ &  7.5 $-$  6.5 & 4276404&.386 &    0&.417 \\
 $ 7.5^{\rm e} -  6.5^{\rm f}$ &  8.5 $-$  7.5 & 4388268&.502 & $-$0&.117 \\
 $ 7.5^{\rm e} -  6.5^{\rm f}$ &  7.5 $-$  6.5 & 4388330&.480 &    0&.058 \\
 $ 7.5^{\rm e} -  6.5^{\rm f}$ &  6.5 $-$  5.5 & 4388383&.531 &    0&.004 \\
 $ 7.5^{\rm f} -  6.5^{\rm e}$ &  6.5 $-$  5.5 & 4390643&.905 &    0&.072 \\
 $ 7.5^{\rm f} -  6.5^{\rm e}$ &  7.5 $-$  6.5 & 4390695&.621 & $-$0&.274 \\
 $ 7.5^{\rm f} -  6.5^{\rm e}$ &  8.5 $-$  7.5 & 4390756&.112 &    0&.257 \\
 $ 9.5^{\rm e} -  8.5^{\rm f}$ & 10.5 $-$  9.5 & 4623467&.645 & $-$0&.168 \\
 $ 9.5^{\rm e} -  8.5^{\rm f}$ &  9.5 $-$  8.5 & 4623532&.062 & $-$0&.214 \\
 $ 9.5^{\rm e} -  8.5^{\rm f}$ &  8.5 $-$  7.5 & 4623589&.493 & $-$0&.149 \\
 $ 9.5^{\rm f} -  8.5^{\rm e}$ &  8.5 $-$  7.5 & 4626538&.285 & $-$0&.095 \\
 $ 9.5^{\rm f} -  8.5^{\rm e}$ &  9.5 $-$  8.5 & 4626588&.085 &    0&.253 \\
 $ 9.5^{\rm f} -  8.5^{\rm e}$ & 10.5 $-$  9.5 & 4626642&.944 & $-$0&.211 \\
 $10.5^{\rm e} -  9.5^{\rm f}$ & 11.5 $-$ 10.5 & 4744561&.295 & $-$0&.439 \\
 $10.5^{\rm e} -  9.5^{\rm f}$ & 10.5 $-$  9.5 & 4744627&.399 & $-$0&.063 \\
 $10.5^{\rm e} -  9.5^{\rm f}$ &  9.5 $-$  8.5 & 4744686&.692 &    0&.032 \\
 $10.5^{\rm f} -  9.5^{\rm e}$ &  9.5 $-$  8.5 & 4747980&.294 &    0&.194 \\
 $10.5^{\rm f} -  9.5^{\rm e}$ & 10.5 $-$  9.5 & 4748028&.065 & $-$0&.254 \\
 $10.5^{\rm f} -  9.5^{\rm e}$ & 11.5 $-$ 10.5 & 4748081&.730 &    0&.024 \\
\hline \hline
\end{tabular}\\[2pt]
}
\end{center}
$^a$ {\footnotesize
Uncertainty 250~kHz for each fine structure line.}
\end{table}

%%%%%%%%%%%%%%%%%%%%%%%%%%%%%%%%%%%%%%%%%%%%%%%%%%%%%%%%%%%%%%%%%%%%%%%%%%%%%%%%%%%%%
%%%%%%%%%%%%%%%%%%%%%%%%%%%%%%%%%%%%%%%%%%%%%%%%%%%%%%%%%%%%%%%%%%%%%%%%%%%%%%%%%%%%%
%%%%%%%%%%%%%%%%%%%%%%%%%%%%%%%%%%%%%%%%%%%%%%%%%%%%%%%%%%%%%%%%%%%%%%%%%%%%%%%%%%%%%

The N$^{18}$O rotational transitions were found easily based on predictions generated from 
previous work \cite{NO_N-15-O_NO-18_rot_1991,NO_N-15-O_NO-18_rot_1999}. Hyperfine structure 
(HFS) was not resolved in these transitions, and good signal-to-noise ratios (SNR) were 
obtained. Uncertainties of 50~kHz were assigned to these data which are gathered in 
Table~\ref{rotational_transitions}. Combining our new data with the previous ones and taking 
into account the fine structure (FS) splitting in NO and $^{15}$NO \cite{NO_N-15-O_rot_1999}, 
the weaker magnetic dipole transition were observed readily. As can bee seen in 
Fig.~\ref{FS-spectrum}, HFS was resolved in these spectral recordings, and the SNR were 
reasonable. We assigned uniformly 250~kHz as uncertainties to these transition 
frequencies mainly because of the lower SNR, but also because of the larger line 
width caused by pressure broadening. The magnetic dipole transitions are summarized in 
Table~\ref{FS_transitions}.

%%%%%%%%%%%%%%%%%%%%%%%%%%%%%%%%%%%%%%%%%%%%%%%%%%%%%%%%%%%%%%%%%%%%%%%%%%%%%%%%%%%%%
%%%%%  Spectroscopic analysis  %%%%%%%%%%%%%%%%%%%%%%%%%%%%%%%%%%%%%%%%%%%%%%%%%%%%%%
%%%%%%%%%%%%%%%%%%%%%%%%%%%%%%%%%%%%%%%%%%%%%%%%%%%%%%%%%%%%%%%%%%%%%%%%%%%%%%%%%%%%%

\section{Spectroscopic analysis}
\label{analysis}

NO is a stable radical with a regular $^2 \Pi$ ground electronic state, i.e., the 
$^2 \Pi _{1/2}$ spin ladder is at lower energies than the $^2 \Pi _{3/2}$ spin ladder. 
It has a fairly small electric dipole moment of 0.15872\,(2)~D \cite{NO_RF_1970}. 
As a diatomic consisting of two fairly light atoms, its spin-orbit splitting is 
comparatively small ($\sim$3.7~THz) while its rotational constant is with 
$\sim$51~GHz fairly large. As a consequence, its spectrum is close to Hund's case 
(a) at lower rotational quantum numbers, but closer to Hund's case (b) at 
intermediate and higher rotational quantum numbers.

The effective Hamiltonian suitable to fit the rotational spectrum of NO has been described 
rather often, and a rather detailed description can be found in Ref.~\cite{NO_N-15-O_rot_1999}. 
Further discussion on the Hamiltonian of a $^2 \Pi$ molecule in terms of Hund's cases 
(a) and (b) can be found elsewhere \cite{radi-Hamiltonian}. Pickett's {\scriptsize SPCAT} and 
{\scriptsize SPFIT} programs \cite{spfit_1991} were used for prediction and fitting of the 
NO spectra. The programs were intended to be rather general, thus being able to fit 
asymmetric top rotors with spin and vibration-rotation interaction in support of the 
spectral line lists of the Jet Propulsion Laboratory (JPL) \cite{JPL-catalog_1998} and 
Cologne Database for Molecular Spectroscopy (CDMS) \cite{CDMS_1,CDMS_2}. 
Hund's case (b) quantum numbers are employed in {\scriptsize SPCAT} and {\scriptsize SPFIT} 
whereas Hund's case (a) quantum numbers are more common for NO. We follow the latter 
labeling in Fig.~\ref{FS-spectrum} and Tables~\ref{rotational_transitions} and 
\ref{FS_transitions}. Conversion of Hund's case (b) quanta to case (a) or vice versa 
depends on the magnitude of the rotational energy relative to the magnitude of the 
spin-orbit splitting. For $2B(J - 0.5)(J + 0.5) < |A|$, levels with $J + 0.5 = N$ 
correlate with $^2 \Pi _{1/2}$ and levels with $J - 0.5 = N$ correlate with 
$^2 \Pi _{3/2}$; for larger values of $J$, the correlation is reversed. In the case 
of the NO isotopologs, the reversal occurs between $J = 5.5$ and $J = 6.5$.

During the fitting process, we contained a spectroscopic parameter in the fit if it reduced 
the rms error of the fit, as measure of the quality of the fit, by an appreciable amount. 
This meant in most instances that the parameter was determined with great significance, 
meaning its uncertainty in the fit was about one fifth of the magnitude of its value or less. 
Care was also taken to evaluate which parameter reduced the rms error by the greatest amount.

Among the available data of one isotopic species and within one vibrational state, 
we used those, which were most accurate because data with larger uncertainties have 
considerably lower weights in the fit; the weight of a datum in the fit scales 
inversely to the square of the uncertainty. In a few cases, multiple data were used 
if the uncertainties were similar. 
We scrutinized the reported uncertainties in all instances. For the great majority 
of the data, the reported values were employed in the fit. Few transition frequencies 
were omitted from the fit if their residuals in the fits were much larger than 
the reported uncertainties. In few other cases with large residuals, the uncertainties 
were increased. Some uncertainties appeared to be conservative, and they were reduced 
somewhat. Details will be given below.

In order to evaluate N$^{18}$O spectroscopic parameters, we combined our data with the 
lower frequency data from Saleck et al. \cite{NO_N-15-O_NO-18_rot_1991}. Uncertainties 
assigned to the transition frequencies pertaining to the lowest quantum numbers of the 
$^2\Pi_{1/2}$ ladder ($J$ = 1.5$-$0.5 and 2.5$-$1.5) appeared to be too conservative, 
not only for N$^{18}$O \cite{NO_N-15-O_NO-18_rot_1991} in the single isotopolog fit, 
but also in the combined fit and for $^{15}$NO \cite{NO_N-15-O_NO-18_rot_1991}, 
and for N$^{17}$O and $^{15}$N$^{18}$O \cite{NO-17_N-15-O-18_rot_1994}; therefore, 
we reduced the uncertainties somewhat for these transitions. We omitted the data from 
Ref.~\cite{NO_N-15-O_NO-18_rot_1999} because they had slightly to considerably larger 
uncertainties, and the data with only slightly larger uncertainties had residuals 
frequently much larger than the quoted uncertainties.

The initial spectroscopic parameter set consisted of those employed for NO and $^{15}$NO 
\cite{NO_N-15-O_rot_1999}. It is worthwhile mentioning that $\gamma$ was used there 
and in the present fits whereas most other NO parameter sets employed $A_D$ instead. 
$A_D$ and $\gamma$ make essentially indistinguishable contributions in a $^2 \Pi$ 
radical, and the same holds for their vibrational or rotational corrections 
\cite{Veseth_2Pi_1971}; only one of the two parameters can be determined usually. 
One way to resolve the indeterminacy is an isotopic invariant fit \cite{AD_gamma_ii_1977}, 
which will be described for NO in the following part of this section. If at least one 
of the atoms of the molecule has a non-zero spin, the combination of HFS and Zeeman 
effects may allow to disentangle $A_D$ and $\gamma$ \cite{AD_gamma_Zeeman_2002}; 
it turned out that in NO the contributions come almost entirely from $\gamma$, whereas 
$A_D$ dominated the contributions in FO \cite{AD_gamma_Zeeman_2002}. 
Several of the initial parameters in the N$^{18}$O fit were poorly determined 
and were omitted successively from the fit without increasing the rms error much. 
The nuclear spin-rotation parameter $C_{I}$ was retained in the fit despite being 
not determined significantly because its omission increased the rms error by more 
than 5\,\% and because its value was correct within its uncertainty. The final set 
of N$^{18}$O spectroscopic parameters is given in Table~\ref{NO-18-parameters}.

%%%%%%%%%%%%%%%%%%%%%%%%%%%%%%%%%%%%%%%%%%%%%%%%%%%%%%%%%%%%%%%%%%%%%%%%%%%%%%%%%%%%%
%%%%%  Table 3  %%%%%%%%%%%%%%%%%%%%%%%%%%%%%%%%%%%%%%%%%%%%%%%%%%%%%%%%%%%%%%%%%%%%%
%%%%%%%%%%%%%%%%%%%%%%%%%%%%%%%%%%%%%%%%%%%%%%%%%%%%%%%%%%%%%%%%%%%%%%%%%%%%%%%%%%%%%

\begin{table}
  \begin{center}
  \caption{Spectroscopic parameters$^a$ (MHz) of N$^{18}$O from a single isotopolog fit.}
  \label{NO-18-parameters}
\smallskip
{\footnotesize
\renewcommand{\arraystretch}{1.10}
  \begin{tabular}{lr@{}l}
  \hline
Parameter                                           & \multicolumn{2}{c}{Value} \\[1pt]
\hline
$B$                                                 &     48\,211&.775\,59~(114) \\
$-D \times 10^3$                                    &      $-$147&.517\,3~(35)   \\
$H \times 10^9$                                     &          32&.7~(30)        \\
$A$                                                 & 3\,691\,991&.767~(53)      \\
$\gamma$                                            &      $-$184&.205~(41)      \\
$\gamma_D \times 10^3$                              &           1&.420~(110)     \\
$p$                                                 &         332&.201\,8~(70)   \\
$q$                                                 &           2&.536\,0~(18)   \\
$q_D \times 10^6$                                   &          37&.3~(25)        \\
$a$                                                 &          84&.214~(39)      \\
$b_{F}$                                             &          22&.425~(142)     \\
$c$                                                 &       $-$58&.871~(200)     \\
$d$                                                 &         112&.585~(11)      \\
$eQq_{1}$                                           &        $-$1&.837~(27)      \\
$eQq_{2}$                                           &          23&.76~(74)       \\
$C_{I} \times 10^3$                                 &          12&.8~(47)        \\
\hline \hline
\end{tabular}\\[2pt]
}
\end{center}
$^a$ {\footnotesize
Numbers in parentheses are 1\,$\sigma$ uncertainties in units of the 
least significant figures.\\}
\end{table}

%%%%%%%%%%%%%%%%%%%%%%%%%%%%%%%%%%%%%%%%%%%%%%%%%%%%%%%%%%%%%%%%%%%%%%%%%%%%%%%%%%%%%
%%%%%%%%%%%%%%%%%%%%%%%%%%%%%%%%%%%%%%%%%%%%%%%%%%%%%%%%%%%%%%%%%%%%%%%%%%%%%%%%%%%%%
%%%%%%%%%%%%%%%%%%%%%%%%%%%%%%%%%%%%%%%%%%%%%%%%%%%%%%%%%%%%%%%%%%%%%%%%%%%%%%%%%%%%%

The ro-vibrational energy levels of a diatomic molecule AB, such as NO, can be 
represented by the Dunham expression \cite{Dunham_1932}

\begin{equation}
\label{Dunham}
E(\varv, J)/h = \sum_{i,j} Y_{ij}(\varv + 1/2)^i N^j (N + 1)^j,
\end{equation}
where the $Y_{ij}$ are the Dunham parameters. In electronic states different from $\Sigma$ 
states, i.e. in states with orbital angular momentum $\Lambda > 0$, the expansion in 
$N(N + 1)$ is often replaced by an expansion in $N(N + 1) - \Lambda ^2$, see, e.g., 
Ref.~\cite{Dunham_BO_Watson2}. The ground electronic state of NO is $^2\Pi$ ($\Lambda = 1$), 
and the expansion is often carried out in $N(N + 1) - 1$, and this expansion was used here. 
The expansion in $N(N + 1)$ is quite common also, see, e.g., the case of the $^2 \Pi$ 
radical BrO \cite{BrO_rot_2001}.

Several isotopic species of AB can be fit jointly by constraining the $Y_{ij}$ to
\cite{Dunham_BO_Watson2,Dunham_BO_Watson1} 

\begin{equation}
\label{BO}
Y_{ij} = U_{ij} \left( 1 + \frac{m_e \varDelta^{\rm A}_{ij}}{M_{\rm A}} 
 + \frac{m_e \varDelta^{\rm B}_{ij}}{M_{\rm B}}\right) \mu^{-(i+2j)/2} 
\end{equation}
where $U_{ij}$ is isotopic invariant, $m_e$ is the mass of the electron, $\mu$ is the 
reduced mass of AB, $M_{\rm A}$ is the mass of atom A, and $\varDelta^{\rm A}_{ij}$ 
is a BOB term assiciated with atom A. The abbreviation $\delta_{ij}^{\rm A}$ 
is sometimes used for $U_{ij} \mu^{-(i+2j)/2} \varDelta_{ij}^{\rm A} m_e/M_{\rm A}$. 
We need to point out that both $\varDelta_{ij}^{\rm A}$ and $\delta_{ij}^{\rm A}$ 
are defined negatively in some publications. Obviously, $\varDelta_{ij}^{\rm B}$ 
and $\delta_{ij}^{\rm B}$ are defined equivalently.

Rotational and vibrational corrections to the $\Lambda$-doubling, FS, and HFS parameters 
have been expressed analogously as  in Eqs.~(\ref{Dunham}) and (\ref{BO}), the isotopic 
dependences were given explicitly, e.g., in Refs.~\cite{radi-Hamiltonian,BrO_rot_2001}. 
Briefly, the lowest order fine structure parameters $A_{00}$ and $\gamma_{00}$ scale with 
1 and $\mu ^{-1}$, respectively. The $\Lambda$-doubling parameters $p_{00}$ and $q_{00}$ 
scale with $\mu ^{-1}$ and $\mu ^{-2}$, respectively. The electron spin-nuclear spin coupling 
parameters $a_{00}$, $b_{F,00}$, $c_{00}$, and $d_{00}$ all scale with the respective nuclear 
$g$ factor $g_N$. In the case of nitrogen, both $^{14}$N and $^{15}$N have non-zero spins 
($I = 1$ and 0.5, respectively). The $^{15}$N/$^{14}$N $g$ factor ratio is $-$1.4027548 
\cite{moments_1989,magnetic_moments_2012}. 
There is only one oxygen nucleus, $^{17}$O, with non-zero spin of 5/2, so no $g$ factor ratios 
needed to be considered for oxygen substitution. The lowest order quadrupole parameters, 
$eQq_{1,00}$, $eQq_{2,00}$, and $eQq_{S,00}$, all scale with the quadrupole moment $Q$, 
but there is only one nucleaus for each atom with $I \ge 1$, $^{14}$N and $^{17}$O. 
The lowest order nuclear spin-rotation parameters $C_{I,00}$ and $C'_{I,00}$ scale 
with $g_N \mu ^{-1}$.

Isotopic invariant fits were carried out for numerous diatomics, among them BrO 
\cite{BrO_rot_2001}, CdH \cite{CdH_ii_2004,ZnH_CdH_ii_2006}, ZnH \cite{ZnH_CdH_ii_2006}, 
CH$^+$ \cite{CH+_fitting_2010}, and O$_2$ \cite{O2_1Delta_2012,O2_3states_2012}.

The atomic masses were taken from a recent compilation \cite{AME_2012}. It includes recent 
improvements for $^{14}$N \cite{Mass-14N_2004}, $^{18}$O \cite{Mass-18O_2009}, and $^{17}$O 
\cite{Mass-17O_2010}. Among these, the updated $^{18}$O value is the most relevant one 
for high resolution spectroscopy.

The aim of the present study was modeling of the ground state rotational spectra of NO isotopic 
species. However, in order to separate contributions of the breakdown of the Born-Oppenheimer 
approximation to a certain spectroscopic parameter from the frequently larger vibrational 
correction to this parameter, see e.g., Refs.~\cite{CdH_ii_2004,ZnH_CdH_ii_2006}, we needed 
to consider some information on vibrationally excited NO. These were NO $\varv = 1$ 
$\Lambda$-doubling data \cite{NO_v=0+1_IR-RF-DR_1977,NO_v=0+1_IR-RF-DR_1981} along 
with heterodyne $\varv$ = 1$-$0 IR transition frequencies 
\cite{NO_heterdyne-IR_1986,NO_heterdyne-IR_1996}. A subsequent study will consider the 
extensive available IR data. The study should include not only experimental transition 
frequencies with appropriate uncertainties, but also intensity information from experimental 
measurements as well as from empirical and theoretical modeling \cite{NO_EDMF_2014}.

We started the combined analysis by determining spectroscopic parameters for the 
ground vibrational state of the main isotopolog. As in Ref.~\cite{NO_N-15-O_rot_1999}, 
electric and magnetic dipole transitions in the THz region were taken from that work, 
and RF and MW $\Lambda$-doubling data were taken from 
Refs.~\cite{NO_N-15-O_RF_1972,NO_RF_1970,NO_RF_1976,NO_v=0+1_IR-RF-DR_1977,NO_v=0+1_IR-RF-DR_1981}. 
We used also unpublished data from Pickett et al. \cite{NO_rot_1979} which had been 
used in prior analyses \cite{NO_rot_1980,NO_N-15-O_NO-18_rot_1991,NO_heterdyne-IR_1996}. 
These data were not only of similar accuracy as the THz data \cite{NO_N-15-O_rot_1999}, 
but also extended from the lowest $J = 1.5 - 0.5$ up to $J = 4.5 - 3.5$. In contrast, 
the THz data started at $J = 3.5 - 2.5$ and $J = 4.5 - 3.5$ for the $^2\Pi_{1/2}$ and 
$^2\Pi_{3/2}$ spin components, respectively \cite{NO_N-15-O_rot_1999}. 
One transition frequency from Ref.~\cite{NO_rot_1979}, ($350989.583 \pm 0.020$)~kHz, was 
omitted from the fit because the residual was about five times the uncertainty. The three 
transition frequencies from Ref.~\cite{NO_RF_1976} with reported uncertaities of 1, 1, 
and 2~kHz, respectively, were assigned uncertainties of 3~kHz in accordance with the residuals. 
Ground and excited state $\Lambda$-doubling data from Ref.~\cite{NO_v=0+1_IR-RF-DR_1977} 
appeared to be judged too conservatively with uncertainties of 20 or 25~kHz. They were 
reproduced to within 5~kHz for the most part even with the reported uncertainties. 
Thus we used 5~kHz as uncertainty for each of these lines. Approximately 30~kHz were 
reported as uncertainties for the ground and excited state $\Lambda$-doubling data from 
Ref.~\cite{NO_v=0+1_IR-RF-DR_1981}. We assigned 20~kHz to all of the $\varv = 1$ data and 
to most of the $\varv = 0$ data; 50~kHz were attributed to the $J$ = 20.5$-$20.5 data 
which appeared to require a larger uncertainty of 30 or 50~kHz.

The choice of spectroscopic parameters for NO in its ground vibrational state was straightforward 
for the most part based on previous analyses \cite{NO_N-15-O_rot_1999,NO_heterdyne-IR_1996}.
The distortion parameter $q_H \approx q_{02}$ had uncertainties in the fits slightly smaller 
than in Ref.~\cite{NO_N-15-O_rot_1999}, but its value was much smaller in magnitude and 
not determined with significance; the value obtained with the final line list was 
($-2.9 \pm 2.0) \times 10^{-10}$~MHz. Hence, it was omitted from the final fit, as 
was done in Ref.~\cite{NO_heterdyne-IR_1996}. No distortion correction was needed for any 
of the HFS parameters with the exception of $d$. It was necessary to include $eQq_S$(N) 
in the fit to reproduce the $\Lambda$-doubling data from Ref.~\cite{NO_N-15-O_RF_1972} well. 
This parameter describes the difference in $eQq_1$ between the $^2\Pi_{1/2}$ and $^2\Pi_{3/2}$ 
spin components. It was used, e.g., in a previous study of BrO \cite{BrO_rot_2001}. 
The approach in the original $\Lambda$-doubling study \cite{NO_N-15-O_RF_1972} was equivalent, 
because two independent parameters $\zeta _1$ and $\zeta _2$ were used to determine the $^{14}$N 
quadrupole coupling within the $^2\Pi_{1/2}$ and $^2\Pi_{3/2}$ substates, respectively.

%%%%%%%%%%%%%%%%%%%%%%%%%%%%%%%%%%%%%%%%%%%%%%%%%%%%%%%%%%%%%%%%%%%%%%%%%%%%%%%%%%%%%
%%%%%  Table 4  %%%%%%%%%%%%%%%%%%%%%%%%%%%%%%%%%%%%%%%%%%%%%%%%%%%%%%%%%%%%%%%%%%%%%
%%%%%%%%%%%%%%%%%%%%%%%%%%%%%%%%%%%%%%%%%%%%%%%%%%%%%%%%%%%%%%%%%%%%%%%%%%%%%%%%%%%%%

\begin{table}
  \begin{center}
  \caption{Spectroscopic parameters$^a$ (MHz) for NO determined from the isotopic invariant fit.}
  \label{spec-parameters}
\smallskip
{\footnotesize
\renewcommand{\arraystretch}{1.20}
  \begin{tabular}{lr@{}l}
  \hline
Parameter                                                             & \multicolumn{2}{c}{Value} \\[1pt]
\hline
$Y_{10}^{\rm eff} \times 10^{-3}$                                     &     56\,240&.216\,66~(14) \\
$U_{01}\mu^{-1}$                                                      &     51\,119&.680\,7~(42)  \\
$U_{01}\mu^{-1}\varDelta _{01}^{\rm N}m_e/M_{\rm N}$                  &        $-$4&.469\,2~(29)  \\
$U_{01}\mu^{-1}\varDelta _{01}^{\rm O}m_e/M_{\rm O}$                  &        $-$4&.027\,2~(27)  \\
$Y_{11}$                                                              &      $-$526&.763\,3~(22)  \\
$U_{02}\mu^{-2} \times 10^3$                                          &      $-$163&.944\,1~(30)  \\
$U_{02}\mu^{-2}\varDelta _{02}^{\rm N}m_e/M_{\rm N} \times 10^3$      &           0&.044\,7~(24)  \\
$Y_{12} \times 10^3$                                                  &        $-$0&.484\,2~(55)  \\
$Y_{03} \times 10^9$                                                  &          37&.940~(114)    \\
$A_{00}^{\rm BO}$                                                     & 3\,695\,104&.22~(65)      \\
$A_{00}^{\rm BO}\varDelta _{00}^{A{\rm ,N}}m_e/M_{\rm N}$             &         204&.98~(26)      \\
$A_{00}^{\rm BO}\varDelta _{00}^{A{\rm ,O}}m_e/M_{\rm O}$             &         167&.83~(38)      \\
$A_{10}$                                                              &   $-$7\,335&.247~(55)     \\
$A_{01}$                                                              &           0&.122\,8~(59)  \\
$\gamma_{00}$                                                         &      $-$193&.40~(21)      \\
$\gamma_{10}$                                                         &           7&.476\,3~(55)  \\
$\gamma_{01} \times 10^3$                                             &           1&.611\,0~(56)  \\
$p_{00}^{\rm BO,eff}$                                                 &         350&.623\,40~(91) \\
$p_{00}^{\rm BO}\varDelta _{00}^{p{\rm ,N}}m_e/M_{\rm N} \times 10^3$ &       $-$17&.11~(93)      \\
$p_{10} \times 10^3$                                                  &      $-$403&.50~(32)      \\
$p_{01} \times 10^6$                                                  &          34&.1~(12)       \\
$q_{00}$                                                              &           2&.844\,711~(39)\\
$q_{10} \times 10^3$                                                  &       $-$44&.282~(65)     \\
$q_{01} \times 10^6$                                                  &          42&.319~(112)    \\
$a_{00}({\rm N})$                                                     &          84&.304\,2~(106) \\
$a_{10}({\rm N}) \times 10^3$                                         &      $-$202&.3~(211)      \\
$b_{F,00}({\rm N})$                                                   &          22&.271~(21)     \\
$b_{F,10}({\rm N}) \times 10^3$                                       &         249&.~(43)        \\
$c_{00}({\rm N})$                                                     &       $-$58&.890\,4~(14)  \\
$d_{00}({\rm N})$                                                     &         112&.619\,47~(132)\\
$d_{10}({\rm N}) \times 10^3$                                         &       $-$30&.3~(27)       \\
$d_{01}({\rm N}) \times 10^6$                                         &         105&.6~(145)      \\
$eQq_{1,00}({\rm N})$                                                 &        $-$1&.898\,6~(32)  \\
$eQq_{1,10}({\rm N}) \times 10^3$                                     &          77&.4~(64)       \\
$eQq_{2,00}({\rm N})$                                                 &          23&.112\,6~(62)  \\
$eQq_{S,00}({\rm N}) \times 10^3$                                     &        $-$6&.89~(83)      \\
$C_{I,00}({\rm N}) \times 10^3$                                       &          12&.293~(27)     \\
$C'_{I,00}({\rm N}) \times 10^3$                                      &           7&.141~(123)    \\
$a_{00}({\rm O})$                                                     &      $-$173&.058\,3~(101) \\
$b_{F,00}({\rm O})$                                                   &       $-$35&.460~(109)    \\
$c_{00}({\rm O})$                                                     &          92&.871~(171)    \\
$d_{00}({\rm O})$                                                     &      $-$206&.121\,6~(70)  \\
$eQq_{1,00}({\rm O})$                                                 &        $-$1&.425~(47)     \\
$eQq_{2,00}({\rm O})$                                                 &       $-$30&.02~(163)     \\
$C_{I,00}({\rm O}) \times 10^3$                                       &       $-$32&.7~(23)       \\
\hline \hline
\end{tabular}\\[2pt]
}
\end{center}
$^a$ {\footnotesize
Numbers in parentheses are 1\,$\sigma$ uncertainties in units of the 
least significant figures.\\}
\end{table}

%%%%%%%%%%%%%%%%%%%%%%%%%%%%%%%%%%%%%%%%%%%%%%%%%%%%%%%%%%%%%%%%%%%%%%%%%%%%%%%%%%%%%
%%%%%%%%%%%%%%%%%%%%%%%%%%%%%%%%%%%%%%%%%%%%%%%%%%%%%%%%%%%%%%%%%%%%%%%%%%%%%%%%%%%%%
%%%%%%%%%%%%%%%%%%%%%%%%%%%%%%%%%%%%%%%%%%%%%%%%%%%%%%%%%%%%%%%%%%%%%%%%%%%%%%%%%%%%%

Vibrational corrections were evaluated next by including heterodyne IR measurements of the 
NO $\varv = 1 - 0$ fundamental band \cite{NO_heterdyne-IR_1986,NO_heterdyne-IR_1996} with 
reported uncertainties in the fit and subsequently the $\Lambda$-doubling transition frequencies 
of NO in its excited vibrational state \cite{NO_v=0+1_IR-RF-DR_1977,NO_v=0+1_IR-RF-DR_1981} 
with uncertainties as described above. The choice of parameters to be included in the fit was 
straightforward for the most part. After inclusion of $d_{10}$ in the fit at most two of the 
three vibrational corrections to $a$, $b_F$, and $c$ could be determined. The best result was 
obtained in the final fits with vibrational corrections to $a$ and $b_F$. Each of these parameters 
led to a modest reduction of the rms error. In case of the $^{14}N$ quadrupole parameters, 
a vibrational correction was only needed for $eQq_{1,00}$.

Inclusion of the $^{15}$NO $\Lambda$-doubling transitions \cite{NO_N-15-O_RF_1972} into 
the fit turned out to be challenging. In fact, whereas the $^2\Pi_{3/2}$ data were fit 
very satisfactorily in the original study \cite{NO_N-15-O_RF_1972}, none of the $^2\Pi_{1/2}$ 
transition frequencies was reproduced there within the uncertainties. They showed deviations 
between more than 5 times to almost 70 times the quoted uncertainties \cite{NO_N-15-O_RF_1972}. 
Interestingly, our initial trial fits of the $^{15}$NO data lead only to a rejection of the 
$\Delta F \ne 0$, $J = 1.5$ transition of the $^2\Pi_{3/2}$ ladder at ($84.589 \pm 0.002$)~MHz 
because of a residual of 13~kHz. In the combined fit, two $^2\Pi_{1/2}$ $\Lambda$-doubling 
transition frequencies with $J = 0.5$ and $\Delta F \ne 0$ were omitted in addition because 
each one deviated from the calculated frequency by about 40 times the reported uncertainty; 
furthermore, two corresponding frequencies with $J = 1.5$ deviated each by about 8~times the 
reported uncertainty and were omitted also. All other transitions were retained in the fit with 
the uncertainties as reported. The rotational \cite{NO_N-15-O_NO-18_rot_1991,NO_N-15-O_rot_1999} 
and FS transition frequencies \cite{NO_N-15-O_rot_1999} were included with the reported 
uncertainties, except for the modifications in the low-$J$ data \cite{NO_N-15-O_NO-18_rot_1991} 
as mentioned above.

The inclusion of the $^{15}$NO data called for BOB parameters for $Y_{01}$ and $A_{00}$ 
to be included in the fit, as was expected. In addition, BOB parameters were required 
for $Y_{02}$ and $p_{00}$ because transition frequencies with very high rotational 
quantum numbers were determined for NO and for $^{15}$NO \cite{NO_N-15-O_rot_1999}.

Subsequently, N$^{18}$O data were used in the fit as described above. Only BOB parameters 
for $Y_{01}$ and $A_{00}$ were needed because the N$^{18}$O data did not reach as high 
quantum numbers as the NO and $^{15}$NO data. Inclusion of $^{15}$N$^{18}$O data did 
not afford any additional parameters. Finally, the N$^{17}$O data were used in the fit. 
Obviously, new parameters were necessary to account for the $^{17}$O HFS splitting; 
no other parameters were introduced to the fit. The final set of spectroscopic parameters 
determined in the fit is given in Table~\ref{spec-parameters}, derived parameters are 
presented in Table~\ref{derived-parameters}.

All input data were reproduced on average to within the uncertainties employed in the fit; 
the rms error of the fit is 0.924. There is some scatter among the various subdata sets, 
but none has residuals on average much larger than 1.0. Among the smallest values are 
$\Lambda$-doubling transitions from Refs.~\cite{NO_v=0+1_IR-RF-DR_1977,NO_v=0+1_IR-RF-DR_1981}. 
The rms error of our N$^{18}$O data is 0.871, slightly better for the pure rotational data 
and slightly worse for the FS data.

%%%%%%%%%%%%%%%%%%%%%%%%%%%%%%%%%%%%%%%%%%%%%%%%%%%%%%%%%%%%%%%%%%%%%%%%%%%%%%%%%%%%%
%%%%%  Table 5  %%%%%%%%%%%%%%%%%%%%%%%%%%%%%%%%%%%%%%%%%%%%%%%%%%%%%%%%%%%%%%%%%%%%%
%%%%%%%%%%%%%%%%%%%%%%%%%%%%%%%%%%%%%%%%%%%%%%%%%%%%%%%%%%%%%%%%%%%%%%%%%%%%%%%%%%%%%

\begin{table}
  \begin{center}
  \caption{Derived spectroscopic parameters$^a$ (MHz) of NO from the isotopic invariant fit.}
  \label{derived-parameters}
\smallskip
{\footnotesize
\renewcommand{\arraystretch}{1.20}
  \begin{tabular}{lr@{}l}
  \hline
Parameter                                                             & \multicolumn{2}{c}{Value} \\[1pt]
\hline
$Y_{01}$                                                              &     51\,111&.184\,2~(11)  \\
$\varDelta _{01}^{\rm N}$$^{b}$                                       &        $-$2&.231\,66~(147)\\
$\varDelta _{01}^{\rm O}$$^{b}$                                       &        $-$2&.296\,99~(156)\\
$Y_{02} \times 10^3$                                                  &      $-$163&.899\,4~(27)  \\
$\varDelta _{02}^{\rm N}$$^{b}$                                       &        $-$6&.96~(37)      \\
$A_{00}$                                                              & 3\,695\,477&.03~(21)      \\
$\varDelta _{00}^{A{\rm ,N}}$$^{b}$                                   &           1&.416\,0~(18)  \\
$\varDelta _{00}^{A{\rm ,O}}$$^{b}$                                   &           1&.324\,3~(30)  \\
$p_{00}$                                                              &         350&.606\,29~(17) \\
$\varDelta _{00}^{p{\rm ,N}}$$^{b}$                                   &        $-$1&.246~(68)     \\
\hline \hline
\end{tabular}\\[2pt]
}
\end{center}
$^a$ {\footnotesize
Numbers in parentheses are 1\,$\sigma$ uncertainties in units of the 
least significant figures.\\}
$^b$ {\footnotesize Unitless.}
\end{table}

%%%%%%%%%%%%%%%%%%%%%%%%%%%%%%%%%%%%%%%%%%%%%%%%%%%%%%%%%%%%%%%%%%%%%%%%%%%%%%%%%%%%%
%%%%%  Discussion  %%%%%%%%%%%%%%%%%%%%%%%%%%%%%%%%%%%%%%%%%%%%%%%%%%%%%%%%%%%%%%%%%%
%%%%%%%%%%%%%%%%%%%%%%%%%%%%%%%%%%%%%%%%%%%%%%%%%%%%%%%%%%%%%%%%%%%%%%%%%%%%%%%%%%%%%

\section{Discussion and conclusion}
\label{Discussion}

We have reproduced extensive rotational data of several NO isotopologs along with 
heterodyne IR data in one isotopic invariant fit. The later inclusion of extensive
rovibrational data may affect some parameters outside the present uncertainties. 
Moreover, additional vibrational and possibly BOB corrections will be required for some 
parameters. The NO $\varv = 1 - 0$ energy difference, in particular, is merely a 
fitting parameter at present.

%%%%%%%%%%%%%%%%%%%%%%%%%%%%%%%%%%%%%%%%%%%%%%%%%%%%%%%%%%%%%%%%%%%%%%%%%%%%%%%%%%%%%
%%%%%  Table 6  %%%%%%%%%%%%%%%%%%%%%%%%%%%%%%%%%%%%%%%%%%%%%%%%%%%%%%%%%%%%%%%%%%%%%
%%%%%%%%%%%%%%%%%%%%%%%%%%%%%%%%%%%%%%%%%%%%%%%%%%%%%%%%%%%%%%%%%%%%%%%%%%%%%%%%%%%%%

\begin{table*}
  \begin{center}
  \caption{Comparison of Born-Oppenheimer breakdown parameters$^a$ $\varDelta _{01}$ and 
           $\varDelta _{02}$ of NO with those of related molecules.}
  \label{BOB-comp}
\smallskip
{\footnotesize
\renewcommand{\arraystretch}{1.20}
  \begin{tabular}{lr@{}lr@{}lr@{}lr@{}lr@{}lr@{}lr@{}lr@{}l}
  \hline
                          & \multicolumn{2}{c}{NO$^b$} & \multicolumn{2}{c}{NS$^c$} & \multicolumn{2}{c}{CO$^d$} & \multicolumn{2}{c}{CS$^e$} & \multicolumn{2}{c}{SiO$^f$} & \multicolumn{2}{c}{SiS$^g$} & \multicolumn{2}{c}{O$_2$$^h$} & \multicolumn{2}{c}{SO$^i$} \\[1pt]
\hline
$\varDelta _{01}^{\rm A}$ &        $-$2&.2317~(15)     &        $-$3&.424~(68)      &        $-$2&.05603~(24)    &        $-$2&.5434~(49)     &        $-$1&.2976~(44)      &        $-$1&.3935~(42)      &        $-$1&.7353~(31)        &        $-$1&.830~(56)      \\
$\varDelta _{01}^{\rm B}$ &        $-$2&.2970~(16)     &        $-$2&.856~(96)      &        $-$2&.09934~(24)    &        $-$2&.3945~(34)     &        $-$2&.0507~(16)      &        $-$1&.8728~(55)      &        $-$1&.7353~(31)        &        $-$2&.700~(24)      \\
$\varDelta _{02}^{\rm A}$ &        $-$6&.96~(37)       &            &               &        $-$6&.3978~(20)     &        $-$4&.9~(57)        &            &                &            &                &            &                  &            &               \\
$\varDelta _{02}^{\rm B}$ &            &               &            &               &            &               &       $-$11&.0~(66)        &            &                &            &                &            &                  &            &               \\
\hline \hline
\end{tabular}\\[2pt]
}
\end{center}
$^a$ {\footnotesize
Numbers in parentheses are 1\,$\sigma$ uncertainties in units of the 
least significant figures.\\}
$^b$ {\footnotesize This work.\\}
$^c$ {\footnotesize Ref.~\cite{NS_isos_rot_1995}.\\}
$^d$ {\footnotesize Ref.~\cite{CO_fit_2012}.\\}
$^e$ {\footnotesize Ref.~\cite{CDMS_2}.\\}
$^f$ {\footnotesize Ref.~\cite{SiO_rot_2013}.\\}
$^g$ {\footnotesize Ref.~\cite{SiS_rot_2007}.\\}
$^h$ {\footnotesize Ref.~\cite{O2_3states_2012}, value for $X ^3\Sigma^-_g$; $a ^1\Delta_g$: $-$1.9144~(56), $b ^1\Sigma^+_g$: $-$2.1333~(74).\\}
$^i$ {\footnotesize Ref.~\cite{SO_isos_rot_1982}.\\}
\end{table*}

%%%%%%%%%%%%%%%%%%%%%%%%%%%%%%%%%%%%%%%%%%%%%%%%%%%%%%%%%%%%%%%%%%%%%%%%%%%%%%%%%%%%%
%%%%%%%%%%%%%%%%%%%%%%%%%%%%%%%%%%%%%%%%%%%%%%%%%%%%%%%%%%%%%%%%%%%%%%%%%%%%%%%%%%%%%

The hyperfine parameters, however, will not be affected by additional IR data because 
of the lower accuracy of the data and because HFS is not resolved in the FTS data. 
Contributions to the interpretation of the NO HFS parameters have been provided 
numerous times, e.g., in 
Refs.~\cite{NO_rot_1956,NO_N-15-O_NO-18_rot_1999,NO_HFS_interpretation_1955,NS_rot_1969}. 
The $^{14}$N HFS parameter of NO and NS are quite similar \cite{NS_rot_1969} with the 
ones of NS being consistently smaller \cite{NS_rot_1969,NS_isos_rot_1995}, hence the 
spin density is smaller at N in NS compared to NO. 

An early RF study yielded a value $eQq_S = -21$~kHz with rather small uncertainty 
of less than 2~kHz \cite{NO_N-15-O_RF_1972}, not in agreement with our value of 
$(-6.89 \pm 0.83)$~kHz, whereas an extended update of that RF study yielded 
$(-9 \pm 8)$~kHz \cite{NO_RF_1976}. The parameter $eQq_S$ was also determined, e.g., 
for BrO \cite{BrO_rot_2001}. If we scale that value of $(-21.817 \pm 0.087)$~MHz with 
the NO/BrO ratios of $A_{00}$ and $eQq_{1,00}$, we obtain a value of $-7.97$~kHz, 
rather close to our NO value. The agreement should not be overinterpreted because 
deriving $eQq_S$ values of ClO and IO analogously from the BrO value yields 
0.96~MHz and 141~MHz compared to experimental values of ($0.313 \pm 0.071$)~MHz 
\cite{ClO_rot_2001} and ($198.17 \pm 0.65$)~MHz \cite{IO_rot_2001}, respectively.

The values $\gamma = -181.15$~MHz and $A_D = 0.169$~MHz, derived from Zeeman spectroscopy 
of NO \cite{AD_gamma_Zeeman_2002}, and our values of $\gamma_{00} = -193.4$~MHz 
or $\gamma = -189.7$~MHz in $\varv = 0$ and $A_D = 0.123$~MHz are in reasonable 
accordance. No uncertainties were quoted in the Zeeman study. But even if uncertainties 
were determined, it would be difficult to judge which values are more reliable. 
A combination of both methods should yield improved values. Both results demonstrate 
that fitting $\gamma$ in rotational or rovibrational spectra of NO isotopologs is 
more appropriate than fitting $A_D$ even though the ratio $A_D/A$ is small 
with respect to $D/B$ if $\gamma$ is omitted from the fit.

An isotopic invariant fit of rotational data of three NO isotopologs was carried out earlier 
\cite{NO_N-15-O_NO-18_rot_1991}. Their values, $\varDelta _{01}^{\rm N} = -2.265~(61)$ 
and $\varDelta _{01}^{\rm O} = -2.345~(34)$, compare very favorably with ours, 
$\varDelta _{01}^{\rm N} = -2.2317~(15)$ and $\varDelta _{01}^{\rm O} = -2.2970~(16)$. 
Our uncertainties are considerably smaller because of extensive additional data for 
NO and $^{15}$NO \cite{NO_N-15-O_rot_1999}, for N$^{18}$O from the present study, and 
data for N$^{17}$O and $^{15}$N$^{18}$O \cite{NO-17_N-15-O-18_rot_1994}.

The available BOB parameters $\varDelta _{01}$ and $\varDelta _{02}$ of NO are compared 
in Table~\ref{BOB-comp} with data of related diatomics. There are three contributions to 
$\varDelta _{01}$ \cite{Dunham_BO_Watson1}: (i) a higher-order semiclassical term that 
originates in the Dunham formalism and is usually very small, (ii) a diabatic (or nonadiabatic) 
term that is proportional to the molecular $g$-value $g_J$, and finally, (iii) an adiabatic 
term that is derived from the experimental $\varDelta _{01}$ value by subtracting the two 
former contributions. The latter contribution appears to depend more on the two atoms in a 
given diatomic molecule than on specifics of this molecule \cite{BOB_4-6_1982}. The second 
contribution is usually the dominant one, and its magnitude is particularly large if there 
exist low-lying electronic states of the same spin-multiplicity in the molecule \cite{Dunham_BO_Watson1}.

The similarity of the NO and CO values should not be overinterpreted because values for, 
e.g., CO$^+$ are quite different: $\varDelta _{01}^{\rm C} = -0.224$~(14) and 
$\varDelta _{01}^{\rm O} = -1.033$~(33) \cite{CO+_rot_2013}. The $\varDelta _{01}$ values 
in Table~\ref{BOB-comp} cover a considerable part of the normal values, and trends are hard 
to detect. The magnitudes of the NS values are comparatively large for a diatomic consisting 
of fairly light atoms, indicative of at least one fairly low lying electronic doublet state 
\cite{Dunham_BO_Watson1,NS_isos_rot_1995}. However, these values are still much smaller in 
magnitude than those of CH$^+$, $\varDelta _{01}^{\rm C} = -7.9749$~(105) and 
$\varDelta _{01}^{\rm H} = -9.2263$~(76) \cite{CH+_fitting_2010}.

The BOB parameters $\varDelta _{0i}$ appear to be usually negative and increasing in magnitude 
with $i$, as in the case of AlH \cite{AlH_IR_1993} or, with more limited values, CH$^+$ 
\cite{CH+_fitting_2010}. As can be see in Table~\ref{BOB-comp}, this is also the case for 
$\varDelta _{02}^{\rm A}$ of CO and NO. In addition, it appears as if the poorly determined 
$\varDelta _{02}$ values of CS are of correct order of magnitude. The remaining BOB parameters 
determined for NO, $\varDelta _{00}^{A{\rm ,N}}$, $\varDelta _{00}^{A{\rm ,O}}$, and 
$\varDelta _{00}^{p{\rm ,N}}$ are all of fairly small magnitude.

Predictions of the rotational spectra of several NO isotopologs will be available in the 
catalog section\footnote{http://www.astro.uni-koeln.de/cdms/} of the CDMS~\cite{CDMS_1,CDMS_2}. 
The line, parameter, and fit files from the isotopic invariant fit are deposited as 
supplementary material. In addition, these files, along with other auxiliary files, 
will be available in the fitting spectra 
section\footnote{http://www.astro.uni-koeln.de/site/vorhersagen/pickett/beispiele/NO/} 
of the CDMS.

%% The Appendices part is started with the command \appendix;
%% appendix sections are then done as normal sections
%% \appendix

%%%%%%%%%%%%%%%%%%%%%%%%%%%%%%%%%%%%%%%%%%%%%%%%%%%%%%%%%%%%%%%%%%%%%%%%%%%%%%%%%%%%%
%%%%%  acknowledgements  %%%%%%%%%%%%%%%%%%%%%%%%%%%%%%%%%%%%%%%%%%%%%%%%%%%%%%%%%%%%
%%%%%%%%%%%%%%%%%%%%%%%%%%%%%%%%%%%%%%%%%%%%%%%%%%%%%%%%%%%%%%%%%%%%%%%%%%%%%%%%%%%%%

\section*{Acknowledgements}

This study was partly supported by a Grant-in-Aid for Scientific Research on Innovative 
Areas by the Ministry of Education, Culture, Sports, Science, and Technology of Japan 
(grant no. 26108507). K.K. is grateful for support of her stay in Cologne by the 
collaborative research grant SFB~956.

\appendix

\section*{Appendix A. Supplementary Material}

Supplementary data for this article are available on ScienceDirect (www.sciencedirect.com) 
and as part of the Ohio State University Molecular Spectroscopy Archives 
(http://library.osu.edu/sites/msa/jmsa\_hp.htm). Supplementary data associated with this 
article can be found, in the online version, at doi: .

%% References
%%
%% Following citation commands can be used in the body text:
%% Usage of \cite is as follows:
%%   \cite{key}         ==>>  [#]
%%   \cite[chap. 2]{key} ==>> [#, chap. 2]
%%

%% References with bibTeX database:

%%%  \bibliographystyle{elsarticle-num}
%%%  \bibliography{<your-bib-database>}

%% Authors are advised to submit their bibtex database files. They are
%% requested to list a bibtex style file in the manuscript if they do
%% not want to use elsarticle-num.bst.

%% References without bibTeX database:

%%%%%%%%%%%%%%%%%%%%%%%%%%%%%%%%%%%%%%%%%%%%%%%%%%%%%%%%%%%%%%%%%%%%%%%%%%%%%%%%%%%%%
%%%%%%%%%%%%%%%%%%%%%%%%%%%%%%%%%%%%%%%%%%%%%%%%%%%%%%%%%%%%%%%%%%%%%%%%%%%%%%%%%%%%%
%%%%%%%%%%%%%%%%%%%%%%%%%%%%%%%%%%%%%%%%%%%%%%%%%%%%%%%%%%%%%%%%%%%%%%%%%%%%%%%%%%%%%

\end{document}